\documentclass[preprint,aps,showpacs]{revtex4}
\usepackage{graphics,graphicx,epsfig}

\begin{document}
\title{Maximal and minimal height distributions of fluctuating interfaces}
\author{T. J. Oliveira${}^{a)}$ and F. D. A. Aar\~ao Reis${}^{b)}$
\footnote{a) Email address: tiagojo@if.uff.br\\
b) Email address: reis@if.uff.br}}
\affiliation{Instituto de F\'\i sica, Universidade Federal
Fluminense, Avenida Litor\^anea s/n, 24210-340 Niter\'oi RJ, Brazil}

\date{\today}

\begin{abstract}
We study numerically the maximal and minimal height distributions (MAHD, MIHD)
of the nonlinear interface growth equations of second and fourth order and of
related lattice models in two dimensions. MAHD and MIHD are different due
to the asymmetry of the local height distribution, so that, in each class,
the sign of the relevant nonlinear term determines which one of two universal
curves is the MAHD and the MIHD. The average maximal and minimal heights scale
as the average roughness, in contrast to Edwards-Wilkinson (EW) growth. All
extreme height distributions, including the EW ones, have tails that cannot be
fit by generalized Gumbel distributions.
\end{abstract}

\maketitle

Extreme value statistics (EVS) has already been 
applied in several fields of science and engineering
\cite{gumbel,bramwell,aplevs}. It has recent
important applications in surface science, e. g. for modeling the
evolution of corrosion damage at time scales not easily accessible to
experiment \cite{engelhardt}. In uncorrelated random variable sets,
the statistics of the $n$th extrema is described by the Gumbel's first
asymptotic distribution \cite{gumbel,fisher} if the probability density
functions (PDF) of those sets decrease faster than a power law. However,
deviations from this statistics are expected in fluctuating interfaces if there
are strong correlation of local heights. In
one dimension, this is the case of Edwards-Wilkinson (EW) interfaces (Brownian
curves) \cite{shapir,comtet} and other Gaussian interface models
\cite{evsgaussian}. On the other hand, fluctuations of other global
quantities in various physical systems follow generalized Gumbel distributions
\cite{bramwellprl,antal2001,lee}, i. e. the first asymptotic distribution with
noninteger $n$ values. This is
explained by the connections between the EVS of correlated variables and sums of
independent variables drawn from exponential PDF \cite{bertin}, which shows that
Gumbel statistics goes far beyond the description of uncorrelated variables
sets.

The interface models where maximal height distributions (MAHD) were previously
calculated \cite{shapir,comtet,evsgaussian} are symmetric with
respect to the average height, consequently MAHD and minimum height
distributions (MIHD) are the same. However, there is a large number of real
interfaces where the up-down symmetry is broken \cite{barabasi}, such as those
described by the nonlinear
growth models of Kardar-Parisi-Zhang (KPZ) \cite{kpz} and of Villain-Lai-Das
Sarma (VLDS) for molecular beam epitaxy \cite{vlds}, which raises the
question whether MAHD and MIHD are
the same in those systems. This is particularly important for two-dimensional
interfaces due to the variety of real growth processes which show  KPZ
\cite{krim,expkpz} and VLDS scaling \cite{expvlds}.
Recent works on persistence in VLDS growth also motivate such study, since
different exponents
for positive and negative height persistence were obtained \cite{constantin}.
Moreover, the above scenario raises additional (and not less important)
questions for the nonlinear models. The first one is
connected to the possibility of fitting their extreme height distributions
(EHD) by generalized Gumbel distributions, similarly to other correlated
systems. The second one is the scaling of the
average maximal height, since EW interfaces showed an unanticipated scaling as
the square of the average roughness \cite{lee}, in
contrast to several one-dimensional interfaces. This is essential to
correlate surface roughness with the extreme events.

The aim of this letter is to address those questions by performing a numerical
study of the MAHD and MIHD in the steady states of the KPZ and VLDS equations
and of various lattice models belonging to those classes in $2+1$ dimensions. We
will show that, for
each growth class, two universal distributions are obtained, which may be a
MAHD or a MIHD of a given model depending on the sign of the coefficient of the
relevant nonlinear term. Combination of data collapse and extrapolation of
amplitude ratios (e. g. skewness and kurtosis) of those distributions are used
to separate systems with coefficients of different signs. In order to
illustrate the drastic effects that asymmetric PDF (i. e. distributions of
local heights) may have on MAHD and MIHD, we will discuss their differences in
a random deposition-erosion
model on an inert flat substrate. We will also show that average maximal and 
minimal heights is all those models scale as the average roughness, as usually
expected, which shows that the EW scaling is an exception \cite{lee}. Finally,
we will show that KPZ, VLDS and EW distributions cannot be fit by generalized
Gumbel distributions.

MAHD and MIHD were calculated for the KPZ equation ${{\partial h}\over{\partial
t}} = \nu_2{\nabla}^2 h + \lambda_2 {\left( \nabla h\right) }^2 + \eta
(\vec{x},t)$ [$\langle \eta\left(\vec{x},t\right) \eta (\vec{x'},t' )
\rangle = D\delta^d (\vec{x}-\vec{x'} ) \delta\left( t-t'\right)$]  in dimension
$d=2$,
with $\nu_2=0.25$, $D=5\times {10}^{-3}$ and $\lambda_2=\sqrt{75}$
($g\equiv {\lambda_2}^2D/{\nu_2}^3 =24$), in discretized boxes with spatial step
$\Delta x=1$, time increment $\Delta t=0.04$ and linear sizes $8\leq L\leq 64$.
A simple Euler integration method \cite{moser} and a
scheme for suppression of instabilities  \cite{dasgupta} were adopted.
We also simulated three discrete KPZ models in sizes $32\leq L\leq 256$: the
restricted solid-on-solid
(RSOS) model \cite{kk}, the ballistic deposition (BD) \cite{vold} and the
etching
model of Mello et al \cite{mello}. From inspection of their growth rules
\cite{hagston} one knows that $\lambda_2 >0$ for BD and the etching model and
$\lambda_2 <0$ for the RSOS model. The VLDS equation
${{\partial h}\over{\partial t}} = -\nu_4{\nabla}^4 h + \lambda_4 {\nabla}^2
{\left( \nabla h\right) }^2 + \eta (\vec{x},t)$ was integrated with $\nu_4=1$,
$\lambda_4=1$, $D=1/2$, and ${\Delta t} =0.01$, using the same methods, in
sizes $8\leq L\leq 32$. We also simulated a generalized conserved RSOS model
(CRSOS) \cite{crsos}, whose original version was proposed in Ref.
\protect\cite{crsosorig}) and which belongs to the VLDS class, in sizes
$16\leq L\leq 128$. The EW
equation (KPZ with $\lambda_2=0$) was integrated with $\nu=1.5$, $D=1/2$ and
$\Delta t=0.01$ in box sizes $8\leq L\leq 64$. For each model and each lattice
size, distributions with at least
${10}^7$ different configurations were obtained to ensure high accuracy, which
is particularly important at their tails. The extremes were calculated
relatively to the average height of each configuration, the minima being
absolute values of the differences from the average.

In Fig. 1a we show the scaled MAHD and MIHD of the KPZ
equation in box size $L=64$. In these plots, $P(m)dm$ is the probability that
the extreme lies in
the range $\left[ m, m+dm\right]$, $x\equiv \left( m-\langle m\rangle
\right) /\sigma$ and $\sigma\equiv {\left( \langle m^2\rangle -
{\langle m\rangle}^2\right)}^{1/2}$. The high accuracy in Fig. 1a allows us to
distinguish those curves (log-linear plots also show
discrepancies in the right tails). Results for smaller box sizes show that the
finite-size effects are negligible, confirming that MAHD and MIDH are actually
different.

Gumbel's first asymptotic distribution used to compare our
data is $g\left( x;n\right)
= \omega \exp{\left( -n\left[ e^{-b\left( x+s\right)} +b\left( x+s\right)
\right]\right)}$,
where $b=\sqrt{\psi '\left( n\right)}$, $s=\left[ \ln{n}-\psi\left( n\right)
\right] /b$ and $\omega =n^nb/\Gamma\left( n\right)$, with
$\Gamma\left( x\right)$ the Gamma function and $\psi\left( x\right) =
{\partial\ln{\Gamma\left( x\right)}}/\partial{x}$ \cite{bramwellprl,lee}.
In Fig. 1b, we show the MAHD of the KPZ equation and the Gumbel curve with the
same skewness
$0.79$, which has $n=1.95$. Although the fit near the peak is
reasonable, there are significant differences in the tails. Data for different
lattice sizes in Fig. 1b show that discrepancies are not consequence of
finite-size effects. Similar
disagreement is observed when we try to fit the MIHD with a Gumbel curve of
skewness $0.65$ ($n=2.75$). However, the right tails of MAHD and MIHD of the KPZ
class tend to simple exponential decays for large $m$, similarly to the
Gumbel curves.

In Fig. 2a we show the scaled MAHD of the
integrated KPZ equation and of BD, and the MIHD of the RSOS model. In Fig. 2b
we show the MIHD of the KPZ equation and of the etching model, and the MAHD of
the RSOS model. There is excellent data collapse in both plots, which confirm
that MAHD (MIHD) of models with $\lambda_2 >0$ are equal to MIHD (MAHD) of
models with $\lambda_2 <0$. This illustrates the possibility of using the MAHD
and MIHD to identify the sign of the coefficient of the nonlinear term in cases
where it is not known a priori. At this point, EVS is superior to the scaling of
the local height distributions (the PDF), which might also reveal the sign of
the nonlinear terms if distortion by huge finite-size effects were not so
frequent \cite{kpz2d}. 

The visual agreement between those distributions and the small finite-size
effects are quantitatively confirmed by estimates of their skewnesses and
kurtosis in various system sizes.
Figs. 3a and 3b show the skewness of the same models of Figs. 2a and 2b,
respectively, as a
function of $1/L^{1/2}$ (BD data were not shown in Figs. 3a and 3b because they
superimpose the etching model data). The small finite-size dependence of the
data for the
KPZ equation, BD and the etching models leads to $S\approx 0.79$ for
$\left( \lambda_2>0\right)$-MAHD and $\left(\lambda_2<0\right)$-MIHD, and
$S\approx 0.65$ for
$\left(\lambda_2>0\right)$-MIHD and $\left(\lambda_2<0\right)$-MAHD.
Surprisingly, the largest finite-size effects are observed in the
RSOS data, which uses to be the best discrete KPZ model for numerical study of
roughness scaling \cite{kpz2d}.

The MAHD and MIHD for the VLDS equation in box size $L=32$ are plotted in
Fig. 4a. Differences in the peaks are tiny, but discrepancies in the tails are
clearly observed. Finite-size effects are also negligible in this case.
The MAHD has skewness $S\approx 0.63$  and the MIHD has $S\approx 0.55$. In
contrast to the KPZ models, those distributions
have Gaussian-shaped right tails [$\exp{\left( -m^2\right)}$]. For this
reason, fits with generalized Gumbel
distributions (which tend to simple exponentials as $m\to\infty$) are not
possible. This is confirmed for MAHD in Fig. 4a by
comparison with Gumbel's curve with $n=2.90$, which has skewness $0.63$.

For the CRSOS model, MAHD and MIHD show significant finite-size effects,
similarly to the RSOS model. However, extrapolation of the skewness of
distributions in finite-size lattices, shown in Fig. 4b, suggest that MAHD and
MIHD of both models are asymptotically the same. This means that $\lambda_4 >0$
for the CRSOS model,
similarly to its one-dimensional original version, which is exactly solvable
\cite{park}.

The difference between MAHD and MIHD can be easily explained in a model of
random deposition and erosion with an inert flat substrate in the
erosion-dominated
regime. Let $q>1/2$ be the probability of single-particle erosion and $1-q$ of
deposition, and assume that erosion is possible only if $h>0$. A steady state
is attained with average height (relatively to the substrate) $\langle
h\rangle = q/\left( 2q-1\right)$ and PDF $P\left( h\right) \propto
\exp{\left( -h/\langle h\rangle \right)}$ (for $q$ close to $1/2$).
Now consider that one measures the extremes in a set of $L$ (independent)
columns. With that PDF, MAHD is given by Gumbel's first asymptotic distribution
with $n=1$. For large $L$, the minimum absolute height is typically at the
substrate, thus fluctuations of the relative minima are dominated by
fluctuations of the average height, which are Gaussian [$\propto
\exp{\left( -m^2/a\sqrt{L}\right)}$ , $a\equiv
\frac{2q-1}{\sqrt{q\left( 1-q\right)}}$], and so it is the MIHD. The difference
between MAHD and
MIHD is easily confirmed by visual inspection of the plots of these functions.
It is related to the highly asymmetric local height distribution of this model
(skewness of PDF is $S_{PDF}=2$), in
contrast to the slight asymmetry of KPZ ($S_{PDF}\approx 0.26$
\cite{kpz2d}) and VLDS ($S_{PDF}\approx 0.20$ \cite{crsos}).

Now we analyze the average values of extremes of KPZ and VLDS interfaces. They
are assumed to scale as $\langle m\rangle \sim L^{\alpha_m}$, while 
the average roughness scales with the roughness exponent $\alpha$. 
We estimate $\alpha_m$ by extrapolation of
effective exponents $\alpha_m\left( L\right) \equiv {
\ln{ \left[{\langle m\rangle} \left( L\right)
/ {\langle m\rangle}\left( L/2\right)\right] }\over \ln{2} }$, as shown in
Figs. 5a (KPZ) and 5b (VLDS). The estimates of $\alpha_m$ are consistent with
the best known estimates of the roughness exponents $\alpha\approx 0.39$ (KPZ)
\cite{kpz2d} and $\alpha\approx 0.67$ (VLDS) \cite{crsos}. Consequently, the
average values of extremes scale as the average roughness in the KPZ and VLDS
classes in $2+1$ dimensions. This contrasts with the scaling as the squared
roughness in the EW class \cite{lee}, which we also confirmed by simulation.

Finally, in Fig. 6 we show the MAHD of the EW equation and the
generalized Gumbel distribution with $n=2.6$, which has the same skewness.
Despite the good agreement in almost three decades of the scaled $P(m)$, the
discrepancy in the tails is clear. Again, data for two box sizes ($L=64$ and
$L=32$) show that this is not caused by finite-size effects nor to low accuracy
of the data. On the other hand, the analytical prediction by Lee \cite{lee} of a
Gaussian-shaped tail [$\sim\exp{\left( -m^2 \right)}$] of the MAHD is confirmed
by the trend of our data for large $m$. Together with the above results for KPZ
and VLDS classes, it shows
that EVS of important interface growth models in two dimensions are not
connected to the EVS of independent variables, despite the wide applicability
of Gumbel statistics to correlated systems.

In summary, we showed that interface growth models with asymmetric local height
distributions have different maximal and minimal height distributions, the most
important examples being the KPZ and the VLDS classes in two dimensions. In each
class, a pair of universal curves may be maximal of minimal height
distributions depending on the sign of the relevant nonlinear term. The average
maximal and minimal heights of KPZ and VLDS models scale as the average
roughness, in contrast to the EW class. All extreme height distributions,
including the EW ones, cannot be fit by generalized Gumbel distributions.
Although most works on statistical properties of interfaces focus on features of
height distributions and/or roughness scaling \cite{barabasi}, recent studies
show that the statistics of global quantities are very useful to characterize
real growth processes \cite{bramwell,moulinet}. The EVS has the same advantages
of roughness distribution scaling for this task, such as weak finite-size
effects, and also reveals the sign of the nonlinear terms. Information on rare
events is also essential in systems where drastic changes in the dynamics occur
if the global minima or maxima attain certain values, such as in corrosion
damage. On the other hand, for some applications (from friction to parallel
computing) the distributions of local extremes may be important, and the
present study certainly motivates additional studies of those quantities
\cite{toroczkai,hivert}.

\acknowledgments

TJO acknowledges support from CNPq and FDAAR acknowledges support from CNPq and
FAPERJ (Brazilian agencies).

%~~~~~~~~~~~~~~~~~~~~~~~~~~~~~~~~~~~~~~~~~~~~~~~~~~~~~~~~~~~~~~~~~~~~~~~~~~~
%~~~~~~~~~~~~~~~~~~~  REFERENCES  ~~~~~~~~~~~~~~~~~~~~~~~~~~~~~~~~~~~~~~~~~~
%~~~~~~~~~~~~~~~~~~~~~~~~~~~~~~~~~~~~~~~~~~~~~~~~~~~~~~~~~~~~~~~~~~~~~~~~~~~

\begin{figure}[!h]
\includegraphics[width=10cm]{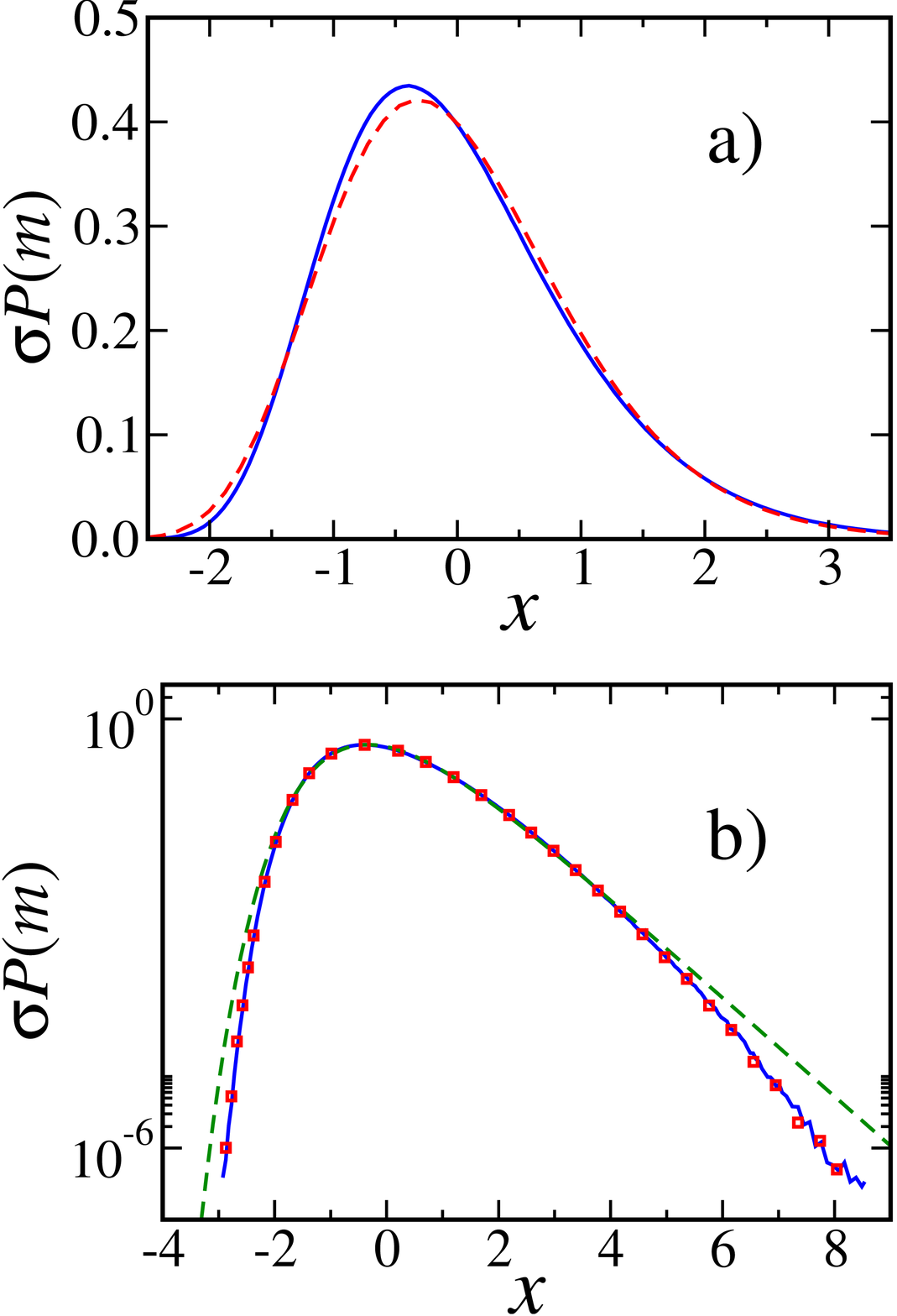}
\caption{(Color online) a) Scaled MAHD (solid curve) and MIHD (dashed curve) of
the KPZ equation in box size $L=64$. b) Scaled MAHD of the KPZ equation (solid
curve for $L=64$, squares for $L=32$) and generalized Gumbel distribution with
$n=1.95$ (dashed curve).}
\label{fig1}
\end{figure}

\begin{figure}[!h]
\includegraphics[width=10cm]{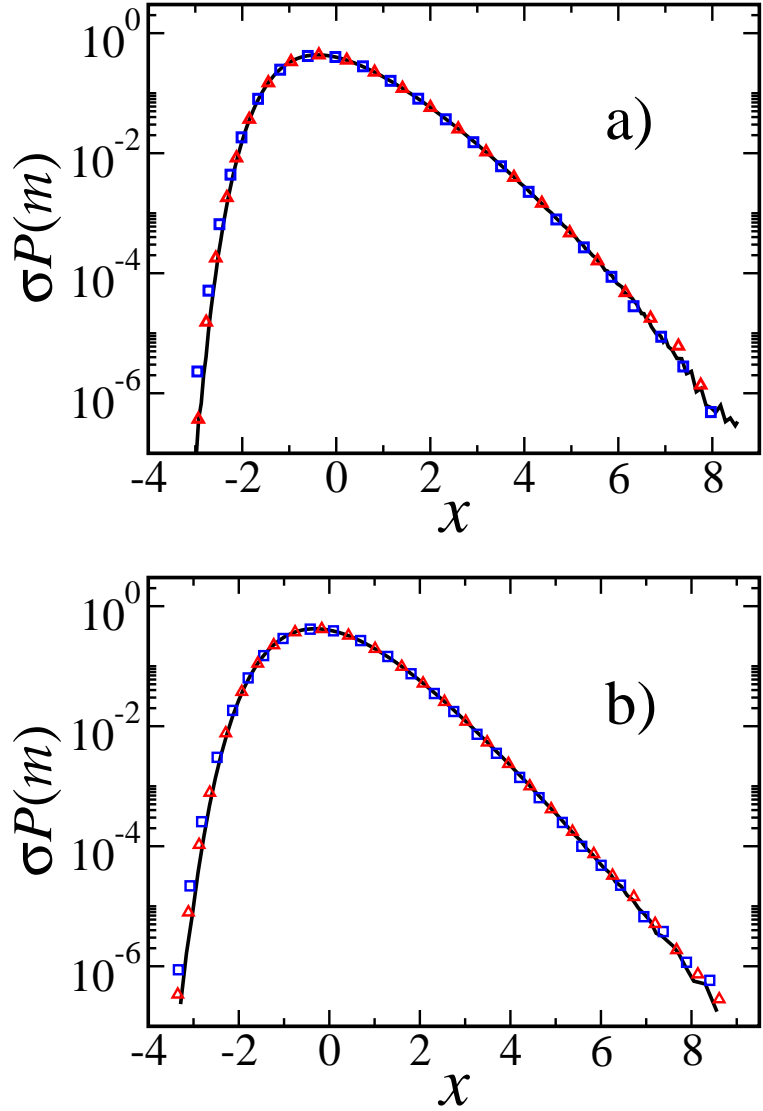}
\caption{(Color online) a) Scaled MAHD of the KPZ equation (solid curve) and BD
(triangles), and MIHD of the RSOS model (squares). b) Scaled MIHD of the KPZ
equation (solid curve) and etching model (triangles), and MAHD of the RSOS
model (squares). For KPZ equation, box size is $L=64$, and for discrete models
$L=256$.}
\label{fig2}
\end{figure}

\begin{figure}[!h]
\includegraphics[width=10cm]{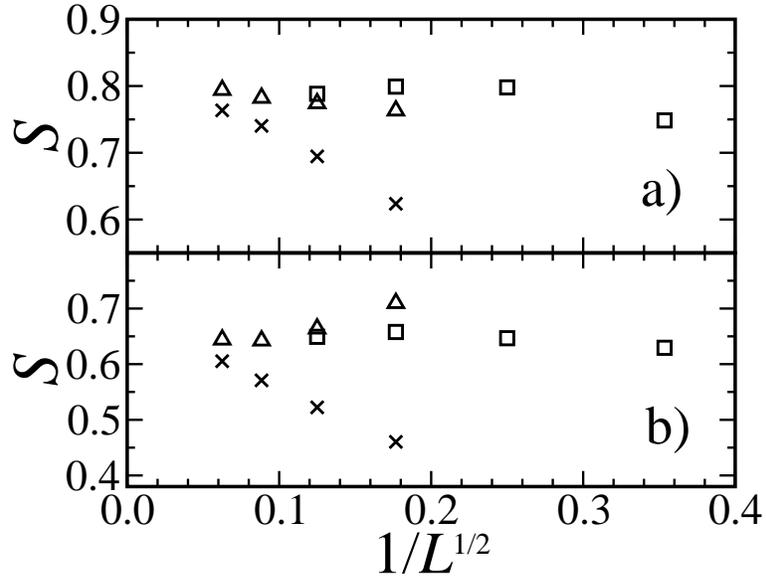}
\caption{a) Finite-size dependence of the skewness $S$ of MAHD of
the KPZ equation (squares) and etching model (triangles), and of MIHD of the
RSOS model (crosses). b) Finite-size dependence of $S$ of MIHD of the KPZ
equation (squares) and etching model (triangles), and of MAHD of the RSOS model
(crosses). In both plots, the variable in the abscissa was chosen to make
clearer the evolution of the data as $L\to\infty$.}
\label{fig3}
\end{figure}

\begin{figure}[!h]
\includegraphics[width=10cm]{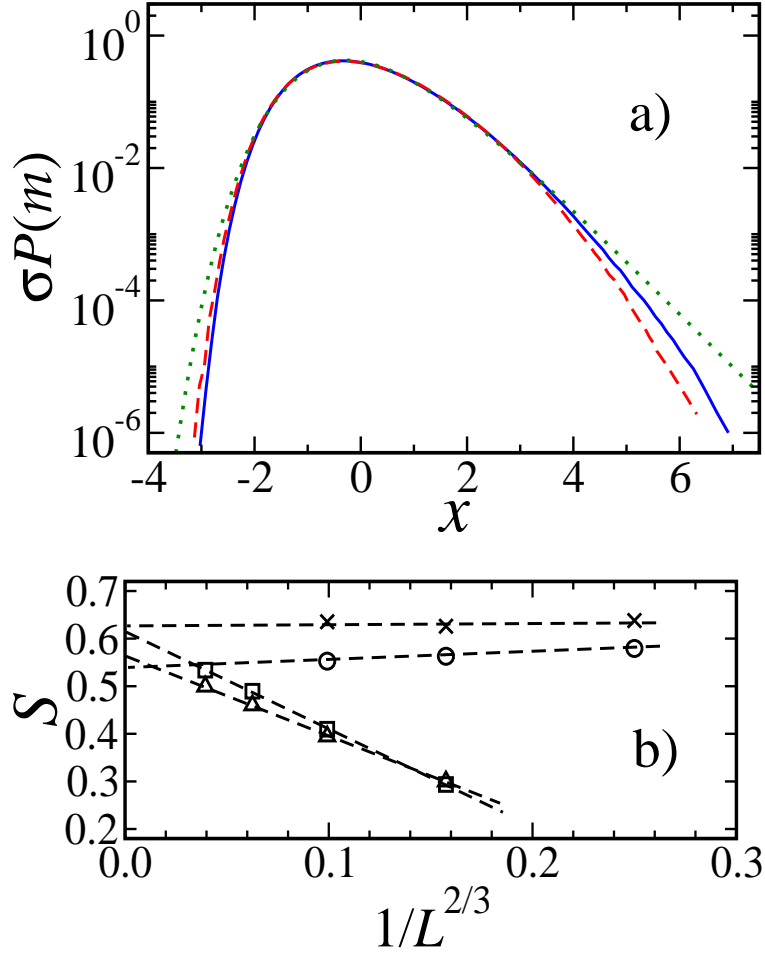}
\caption{(Color online) a) Scaled MAHD (solid curve)  and  MIHD (dashed curve)
of the VLDS equation in $L=32$ and generalized Gumbel distribution with
$n=2.90$ (dotted curve). b) Finite-size dependence of the skewness $S$ of EHD
of the VLDS equation (crosses for MAHD, circles for MIHD) and of the CRSOS
model (squares for MAHD, triangles for MIHD). The variable $1/L^{2/3}$ provides
the best linear fits of the data (dashed lines).}
\label{fig4}
\end{figure}

\begin{figure}[!h]
\includegraphics[width=10cm]{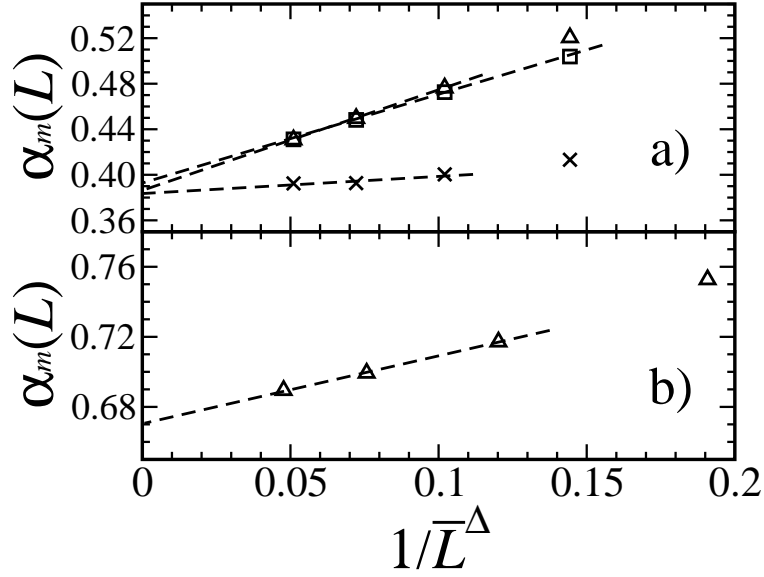}
\caption{Finite-size dependence of effective exponents
$\alpha_m\left( L\right)$ of discrete models: a) KPZ class (BD: crosses;
etching: triangles; RSOS: squares) and b) VLDS class (CRSOS model).
$\overline{L}$ is the average size among $L$ and $L/2$. The variables in the
abscissa provide the best linear fits (dashed lines) with exponents $\Delta
=1/2$ (a) and $\Delta =2/3$ (b).}
\label{fig5}
\end{figure}

\begin{figure}[!h]
\includegraphics[width=10cm]{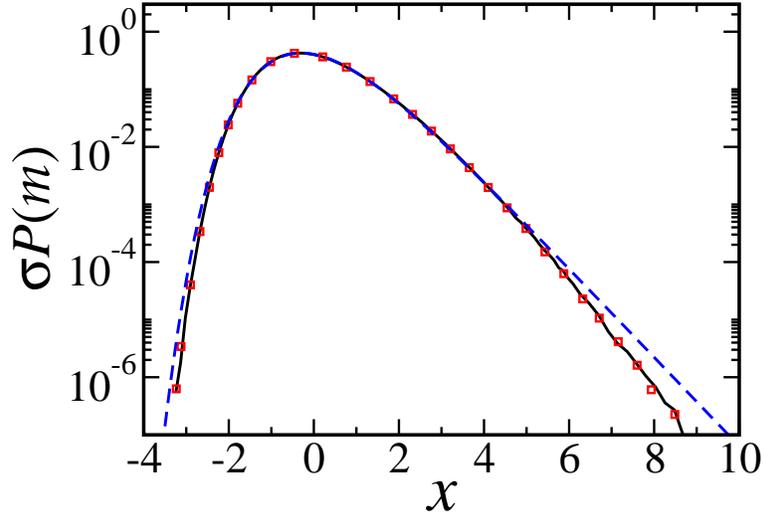}
\caption{(Color online) Scaled MAHD of the EW equation in $L=64$ (solid curve)
and $L=32$ (squares) and the generalized Gumbel distribution with $n=2.6$
(dashed curve).}
\label{fig6}
\end{figure}

\end{document}